\documentclass[aps,prl,twocolumn,letterpaper,10pt]{revtex4}
\usepackage{amssymb}
\usepackage{amsmath}
\usepackage{graphicx}
\usepackage{color}

\newcommand{\Nloc}{N_\text{loc}}

\newcommand{\bNloc}{\bar{N}_\text{loc}}

\newcommand{\tNloc}{N^*}

\begin{document}

\noindent\textbf{Comment on ``Excitons in Molecular Aggregates with L\'evy Disorder: Anomalous Localization and Exchange Broadening of Optical Spectra''\\}

In their Letter~\cite{prl10}, Eisfeld~\textit{et al.}\ predicted exchange broadening and blue-shift of the absorption band as well as a nonuniversal disorder scaling of the localization length of excitons in J-aggregates with L\'evy disorder ($\alpha=\frac{1}{2}$), which they contrasted with the previously analyzed Gaussian and Lorentzian case~\cite{prb09}. The observations were explained by chain segmentation due to outliers, and its interplay with localization of states in effective potential wells created by typical random site energies. We argue that the previously known theory~\cite{prb09} does not break down and thus anticipates the properties of absorption band investigated in the commented work.

\begin{figure}[htbp]
\centering
\includegraphics[scale=0.53]{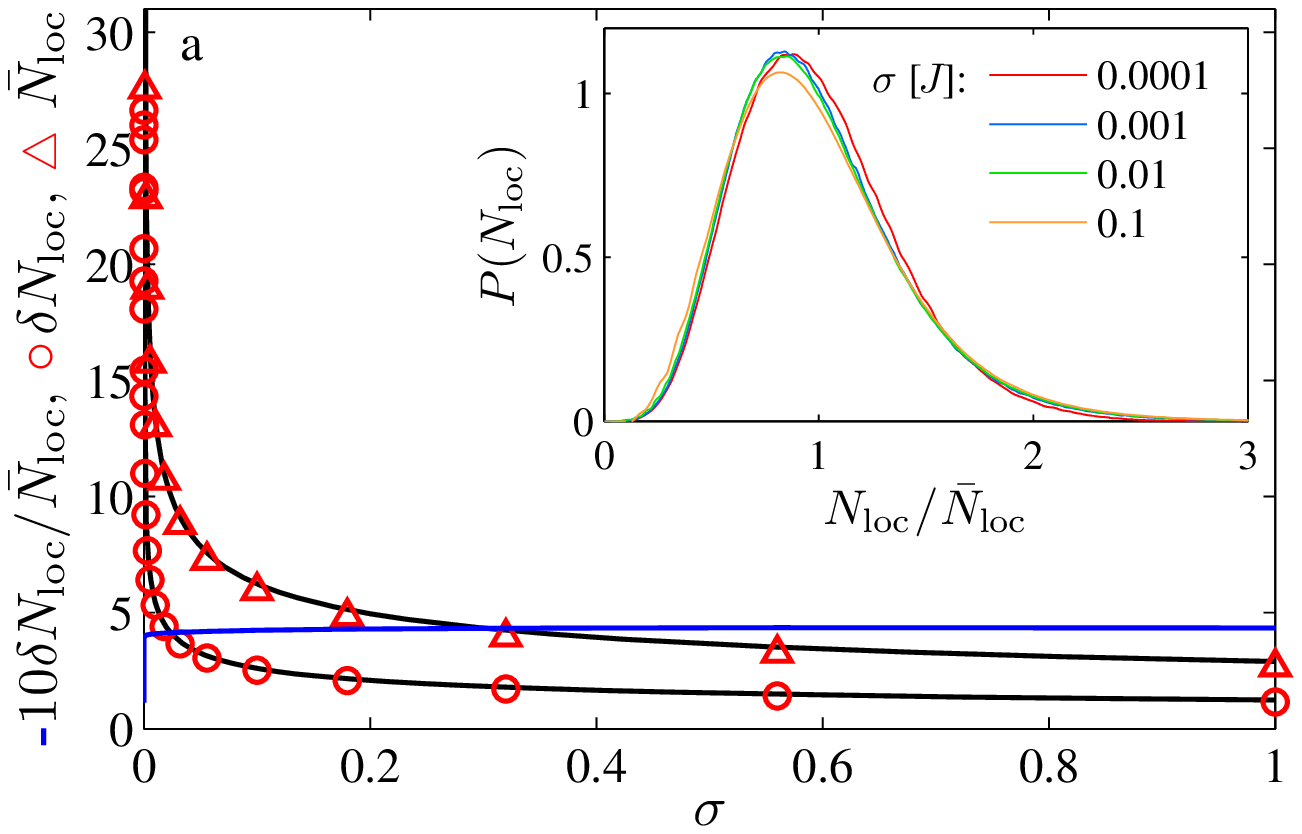}\\
\includegraphics[scale=0.535]{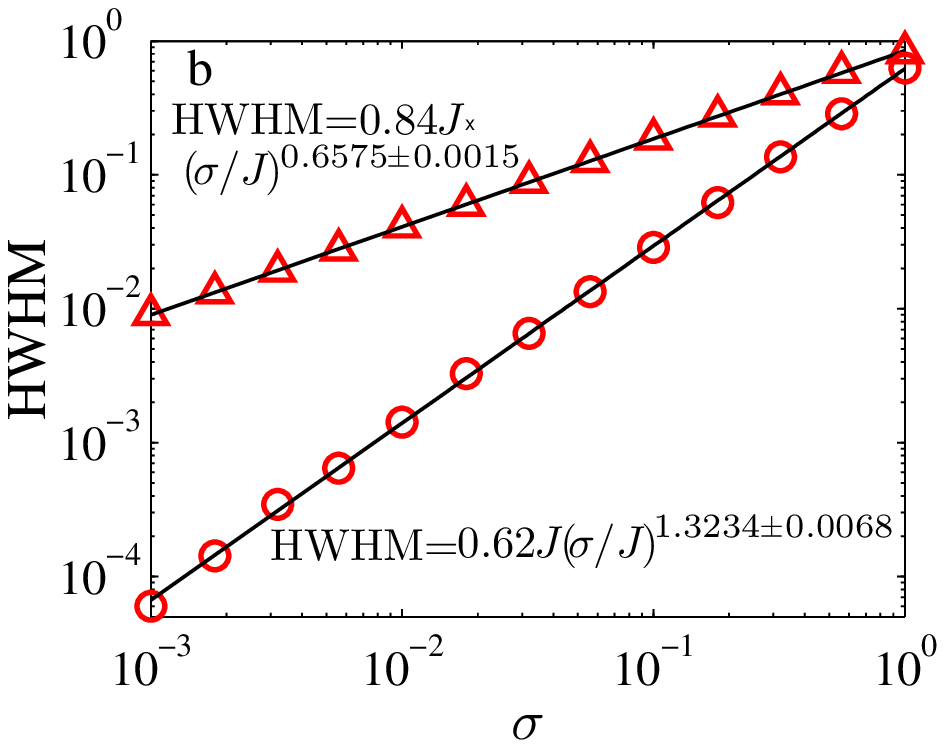}\includegraphics[scale=0.535]{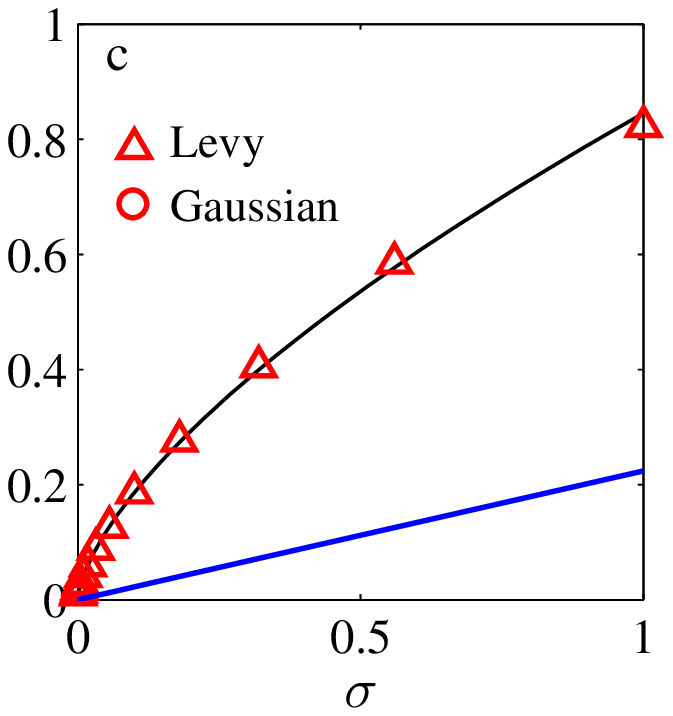}
\caption{a) $\Nloc$ scaling in a chain with L\'evy disorder. Inset: renormalized $\Nloc$ distributions for different $\sigma$ values. b) Power law fits for L\'evy and Gaussian disorder; logarithmic scales should be used for correct results. c) Absorption band vs.\ bare disorder HWHM for L\'evy distribution.}%
\label{fig:fig1}%
\end{figure}

The nonuniversality of the $\Nloc$ distribution shown in Fig.\,3 of~\cite{prl10} does not arise from the properties of L\'evy disorder, but it is an artifact of considering a constant range of energies $\epsilon \in [-2.1,-1.9]J$ for comparison under different scale parameters $\sigma$, instead of an interval adjusting to disorder-induced scaling and shifts of absorption band, as previously done for other disorder types (i.e.~$\tilde{\epsilon} \in [-0.1,0]$)~\cite{prb09}. A $\sigma$-independent energy range implies that for low $\sigma$ values a too wide variety of states is included in the calculation. Consequently, the standard deviation $\delta \Nloc$ disproportionately increases and the ratio $\delta \Nloc/\bNloc$ is no longer constant. Using the correct scaling, $\epsilon = \epsilon_b + (5\tilde{\epsilon} +0.23)\sigma^{2/3}$, we have obtained universal scaling of the $\Nloc$ distribution with $\bNloc\sim(J/\sigma)^{0.333}$ and $\delta\Nloc\sim(J/\sigma)^{0.324}$ (Fig.\,\ref{fig:fig1}a). It breaks down only in the limit of small and large $\sigma$ due to the convergence towards $\Nloc=134$ (unperturbed eigenstates) and $\Nloc=1$, respectively. We have also recalculated the scaling of absorption band HWHM, particularly ensuring good accuracy for low $\sigma$ values, where the band becomes very narrow (Fig.\,1b). The obtained exponent $0.657$, in contrast to 0.6 in the Letter, is in agreement with the theory in~\cite{prb09}. We also note that in Fig.\,2 of~\cite{prl10} the HWHM should be compared not with $\sigma$ but with bare disorder HWHM, equal $\sqrt{2\ln2}\sigma$ for Gaussian and $0.224 \sigma$ for L\'evy distribution, in the latter case revealing the exchange broadening for all $\sigma$ values considered (Fig.\,1c)---not just $[0,0.6]J$.

The above scaling relations result from the Generalized Central Limit Theorem, $\sigma^\ast=\sigma\Nloc^{(1-\alpha)/\alpha}$, and thus work for any $\alpha$. The formula for the site-averaged disorder strength carries the information about the type of its distribution (including heavy-tailedness for $\alpha<2$). The scaling $\tNloc\sim(J/\sigma)^{\alpha/(1+\alpha)}$ results from the equilibrium between its typical value experienced by the absorption band states and the typical energetic cost of localizing them. From the universality of $\Nloc$ distribution follows $\bNloc \sim \tNloc$; together with the above relations, the universality explains also the scaling $\text{HWHM}\sim\sigma(\tNloc)^{(1-\alpha)/\alpha}$. On the other hand, segmentation (already present in Lorentzian disorder~\cite{prb09}) and localization in potential wells are just microscopic mechanisms realizing this statistical theory. The segmentation produces additional peaks in the absorption spectrum, predicted by the Letter, which for sufficiently low $\alpha$ can destroy the HWHM scaling already in the intermediate disorder regime.

AO-C acknowledges support from the EPSRC and AW thanks UCL for hospitality.

{\small Agnieszka Werpachowska (\texttt{a.m.werpachowska@gmail.com}) and Alexandra Olaya-Castro, Dept of Physics and Astronomy, University College London, Gower Street, London WC1E 6BT, United Kingdom. \underline{Authors' contributions}: AW spotting the errors in the Letter, theory and numerics, preparing the Comment, Supplement and responses to Authors and Referees; AOC suggesting reading the Letter, reading the materials.}


\end{document}